\documentclass[aip,
nofootinbib,
rsi,%
reprint,% single-spaced, dense output
twocolumn,
longbibliography,
author-numerical%
]{revtex4-2}

\usepackage{graphicx}
\usepackage{textcomp}

\usepackage{amsfonts}
\usepackage{amssymb}
\usepackage{amsmath}
\usepackage{color}
\usepackage{changes}

\newcommand{\pdr}[2]{\dfrac{\partial {#1}}{\partial {#2}}}

\newcommand{\pdra}[2]{{\partial  {#1}}/{\partial {#2}}}

\newcommand{\tx}{\tilde{x}}

\newcommand{\tJ}{\tilde{J}}

\newcommand{\tN}{\tilde{N}}

\newcommand{\tp}{\tilde{p}}

\newcommand{\Kw}{Q}

\newcommand{\yh}{y_c}
\newcommand{\ph}{\tp_c}

\newcommand{\cref}{c_{ref}}

\newcommand{\lnl}[1]{\ln\left(#1\right)}

\newcommand{\etal}{et al.{ }}

\begin{document}

\sf

\title{Dusty-gas model conservation law and approximate analytical solutions
      for H$_2$--H$_2$O transport in the SOFC anode support layer
      }

\author{Andrei Kulikovsky}
\thanks{ECS member}
\email{A.Kulikovsky@fz-juelich.de}

\affiliation{Forschungszentrum J\"ulich GmbH           \\
    Theory and Computation of Energy Materials (IET--3)   \\
    Institute of Energy and Climate Research,              \\
    D--52425 J\"ulich, Germany
}

% \altaffiliation[Also at: ]{Lomonosov Moscow State University,
%     Research Computing Center, 119991 Moscow, Russia}

% \altaffiliation[Also at: ]{Lomonosov Moscow State University,
%     Research Computing Center, 119991 Moscow, Russia}

\date{\today}

\begin{abstract}
A complete Dusty-Gas Model for the H$_2$--H$_2$O mixture in the anode transport layer
of the anode-supported SOFC is considered. An exact conservation law relating the total
pressure and the hydrogen molar fraction at any point in the anode to their values in
the anode channel is derived. Using this conservation law, approximate analytical solutions
for the hydrogen molar fraction and total pressure in the anode transport layer are obtained.
The solutions can be used to calculate the concentration overpotential.
\end{abstract}

\keywords{Dusty-Gas Model, SOFC anode, two--component mixture, analytical solution}

\maketitle

\section{Introduction}

Anode--supported Solid Oxide Fuel Cells (SOFCs) employ a thick, on the order of 1 mm,
porous anode support layer (ASL). The role of the ASL is threefold: (i) it provides mechanical
stability to the SOFC sandwich, (ii) it serves as a transport layer for gases, and (iii)
it transports electrons from the active layer  to the external circuit.

Mass transport through porous media is a classic problem in chemical engineering~\cite{Mason_83}.
The huge surface area of pores dramatically increases the rate of surface--activated chemical
or electrochemical reactions. In some devices, part of the porous domain is used for diffusive transport
of gaseous or liquid components to/from the reaction zone. An important example is
the anode--supported SOFC, where the porous ASL provides transport
of hydrogen to and water vapor from the thin reaction zone located near the electrolyte~\cite{McEvoy_03}.

\begin{figure}
    \begin{center}
        \includegraphics[scale=1]{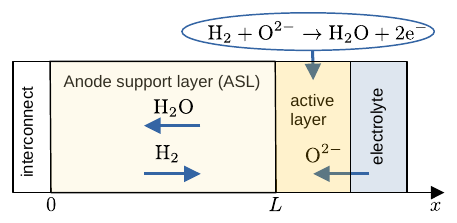}
        \caption{Schematic of anode--supported SOFC anode. The sketch is strongly
           not to scale: the active layer thickness is two orders of magnitude
           smaller than the ASL thickness.
        }
        \label{fig:scheme}
    \end{center}
\end{figure}

Schematic of the SOFC anode is shown in Figure~\ref{fig:scheme}. Hydrogen is supplied through the
channel in the interconnect to the porous ASL, and finally to the active layer, where
the electrochemical conversion of H$_2$ runs
\begin{equation}
   \text{H}_2 + \text{O}^{2-} \to \text{H$_2$O} + 2\rm{e}^{-}
   \label{eq:areac}
\end{equation}
The oxygen ions O$^{2-}$ enter the active layer from the cathode side through the electrolyte.
The product water is removed through the ASL to the channel / interconnect.
The ASL thus supports two opposite fluxes of hydrogen and water vapor.

It has been agreed that the most general description of multicomponent gas transport
in porous media is provided by the Dusty--Gas Model (DGM)~\cite{Mason_83,Zhu_08,Fu_15}.
The DGM equation for the molar fraction $y_k$ of the $k$th component of a gaseous mixture
is (Zhu and Kee~\cite{Zhu_08}):
\begin{multline}
   \sum_{i\neq k}\dfrac{y_i N_k - y_k N_i}{D_{ik}} + \dfrac{N_k}{D_{K,k}} \\
       = -\dfrac{1}{RT}\left(\pdr{(y_k p)}{x} + \dfrac{y_k}{D_{K,k}}\dfrac{p B_0}{\mu}\pdr{p}{x}\right)
      \label{eq:dgm_base}
\end{multline}
or, collecting the terms with the pressure gradient,
\begin{multline}
   \sum_{i\neq k}\dfrac{y_i N_k - y_k N_i}{D_{ik}} + \dfrac{N_k}{D_{K,k}} = {} - \dfrac{p}{RT}\pdr{y_k}{x} \\
      - \dfrac{1}{RT}\left(y_k + \dfrac{y_k}{D_{K,k}}\dfrac{p B_0 }{\mu}\right)\pdr{p}{x}
      \label{eq:dgm_base1}
\end{multline}
Here
$N_k$ is the molar flux of the $k$th component,
$p$ the pressure,
$D_{i,k}$ the effective binary molecular diffusion coefficient,
$D_{K,k}$ the effective Knudsen diffusion coefficient
$B_0$ the hydraulic permeability of the porous media,
$\mu$ the mixture viscosity.
The DGM takes into account the inter--diffusion Stefan--Maxwell fluxes (the first
term on the left side), the Knudsen diffusion in smaller pores (the second term on the left side), and
the flux due to the pressure gradient (the last term in Eq.\eqref{eq:dgm_base1}).

The pressure gradient term introduces considerable complexity to the analysis
and in many works this term has been neglected~\cite{Lehnert_00,Bao_07,Shi_08,Cayan_09,Haeffelin_13,Yang_16}.
This simplifies the DGM, Eq.\eqref{eq:dgm_base1}, to the Stefan--Maxwell--Knudsen Model (SMKM):
\begin{equation}
   \sum_{i\neq k}\dfrac{y_i N_k - y_k N_i}{D_{ik}} + \dfrac{N_k}{D_{K,k}} = {} - \dfrac{p}{RT}\pdr{y_k}{x}
   \label{eq:SMK_base}
\end{equation}
An exact consequence of SMKM is the Graham's law~\cite{Krishna_97}
\begin{equation}
   \sum_k N_k\sqrt{M_k} = 0
   \label{eq:Gra}
\end{equation}
Eq.\eqref{eq:Gra} is obtained by summing Eq.\eqref{eq:SMK_base} over $k$, taking into account
that the Stefan--Maxwell terms cancel, the sum $\sum_k y_k = 1$, and $D_{K,k} \sim 1/\sqrt{M_k}$,
where $M_k$ is the molecular
weight of the $k$th component. However, in the SOFC anode, the water and hydrogen
fluxes are related by the stoichiometric requirement  $N_w = - N_h$, which follows
from Eq.\eqref{eq:areac}. The relation $N_w = - N_h$ is provided by the pressure gradient,
which is missing in SMKM.

% To workaround this problem, SMK
% equation for hydrogen together with the stoichiometric relation for the fluxes,
% instead of Eq.\eqref{eq:Gra} was used (see~\cite{Bertei_15} for discussion).
% Obviously, this approach violates the SMK equation for the second (water) component.

Several works have demonstrated the importance of the pressure gradient
term~\cite{Zhu_05,Kookos_12,Bertei_15} (note a typo in Eq.(35a) of Ref.\cite{Kookos_12}).
Fu \etal\cite{Fu_15} have shown that the inequality of the Knudsen diffusion
coefficients of hydrogen and water creates a pressure gradient in the porous media.
Indeed, in a two--component system, if $M_1$ = $M_2$, the Graham's law, Eq.\eqref{eq:Gra},
reduces to the required SOFC anode stoichiometry relation $N_1 + N_2 = 0$. Thus, if the molecular
weights of water $M_w$ and hydrogen $M_h$ were equal, the SMKM would be equivalent to the DGM.
In fact, however, $\sqrt{M_w/M_h} = 3$, which makes the things more complicated.
Bertei and Nicolella \cite{Bertei_15} discussed this and several other inconsistencies
in models for porous electrodes.

In this work, a complete two--component DGM for the H$_2$--H$_2$O transport
in the SOFC ASL is considered. An exact conservation law  is derived
that relates the total pressure and hydrogen molar fraction at any point in the anode to their
values in the channel. The conservation law is used to construct approximate analytical
solutions to the DGM equations. Comparison with numerical results shows good
accuracy of the analytical formulae.

\section{Model}

\subsection{Two--component DGM equations}

The equation for the hydrogen molar fraction $y \equiv y_h$ in the two--com\-ponent mixture of H$_2$--H$_2$O is
obtained from Eq.\eqref{eq:dgm_base} taking into account that $y_w = 1 - y$ and $N_w = - N_h$:
\begin{equation}
   \pdr{(yp)}{x} + \dfrac{y p B_0 }{D_{K,h}\,\mu}\pdr{p}{x} = - \dfrac{RT}{D_{K,h}}\left(1 + K\right) N_h
      \label{eq:dgm_hy}
\end{equation}
where the subscripts $w$ and $h$ denote water and hydrogen, respectively.
The equation for the total pressure is obtained by summing Eqs.\eqref{eq:dgm_base1}.
It is easy to verify that for the H$_2$--H$_2$O mixture
$\sum_k y_k/D_{K,k} =  (\Kw + y (1-\Kw)) / D_{K,h}$.
Taking into account that the Stefan--Maxwell terms cancel and $\sum_k y_k =1$,
we get
\begin{multline}
   \left(1 + \bigl(\Kw + y\bigl(1 - \Kw\bigr) \bigr)\dfrac{p B_0}{D_{K,h}\,\mu}\right)\pdr{p}{x} \\
      = - \dfrac{RT}{D_{K,h}}\left(1 - \Kw\right) N_h
   \label{eq:dgm_sum}
\end{multline}
Here,
\begin{equation}
   K = \dfrac{D_{K,h}}{D_m}, \quad \Kw = \dfrac{D_{K,h}}{D_{K,w}} = \sqrt{\dfrac{M_w}{M_h}} = 3.
   \label{eq:K}
\end{equation}
The factor $p B_0/(D_{K,h}\,\mu)$ in Eqs.\eqref{eq:dgm_hy}, \eqref{eq:dgm_sum}
 suggests a suitable characteristic scale
for pressure. Introducing dimensionless variables
\begin{equation}
   \tx = \dfrac{x}{L}, \quad \tp = \dfrac{p}{p_*}, \quad p_* = \dfrac{\mu D_{K,h}}{B_0}
   \label{eq:dless}
\end{equation}
and setting $\Kw = 3$,  Eqs.\eqref{eq:dgm_hy}, \eqref{eq:dgm_sum} transform to
\begin{align}
   & \pdr{(y\tp)}{\tx} + y\tp\pdr{\tp}{\tx} ={} - (1 + K)\tN_h
      \label{eq:dgm_hyd} \\
   & \bigl(1 + \tp\,(3 - 2y)\bigr)\pdr{\tp}{\tx} = 2\tN_h
   \label{eq:dgm_sumd}
\end{align}
where the dimensionless hydrogen molar flux is
\begin{equation}
   \tN_h = \dfrac{N_h}{N_*}, \quad N_* = \dfrac{\mu D_{K,h}^2}{R T L B_0}
   \label{eq:tN}
\end{equation}

\subsection{Mass transport equations}

Hydrogen mass conservation in the ASL prescribes that
\begin{equation}
   \pdr{\tN_h}{\tx} = 0.
   \label{eq:mass}
\end{equation}
Differentiating Eqs.\eqref{eq:dgm_hyd}, \eqref{eq:dgm_sumd} over $\tx$, we thus get
\begin{equation}
  \pdr{}{\tx}\left(\pdr{(y\tp)}{\tx} + y\tp\pdr{\tp}{\tx} \right) = 0
   \label{eq:dgm_hyd2}
\end{equation}
\begin{equation}
   \pdr{}{\tx}\left(\bigl(1 + \tp\,(3 - 2y)\bigr)\pdr{\tp}{\tx}\right) = 0
   \label{eq:dgm_sumd3}
\end{equation}
It is worth noting  that the system of Eqs.\eqref{eq:dgm_hyd2}, \eqref{eq:dgm_sumd3}
does not contain parameters.
Eqs.\eqref{eq:dgm_hyd2}, \eqref{eq:dgm_sumd3} form a system of two second--order
nonlinear elliptic equations for $y$ and $\tp$. Eqs.\eqref{eq:dgm_hyd}, \eqref{eq:dgm_sumd} are the
first integral of Eqs.\eqref{eq:dgm_hyd2}, \eqref{eq:dgm_sumd3}
and below, Eqs.\eqref{eq:dgm_hyd}, \eqref{eq:dgm_sumd} will be analyzed further.
% Specific parameters of the system appear only in the boundary conditions.

\subsection{Boundary (initial) conditions}

At the channel/ASL interface $y =  \yh$, $\tp = \ph$. where the subscript $c$
marks the values in the channel. The reaction stoichiometry prescribes
that the hydrogen flux $N_h = J/(2F)$, or $\tN =\tJ$ (see below).
Setting in Eq.\eqref{eq:dgm_hyd} $\tx=0$
and expanding the derivative in the first term, we get
\begin{equation}
     \left. \ph\pdr{y}{\tx}\right|_{\tx=0}
        =  -  \yh\bigl(1 + \ph\bigr)\left.\pdr{\tp}{\tx}\right|_{\tx=0} - (1 + K)\tJ
    \label{eq:dydx0}
\end{equation}
where
\begin{equation}
   \tJ = \dfrac{J}{J_*}, \quad J_* = 2 F N_* = \dfrac{2 F \mu D_{K,h}^2}{R T L B_0}
   \label{eq:tJ}
\end{equation}
Quite similarly, setting $\tx=0$ in Eq.\eqref{eq:dgm_sumd}, we find
\begin{equation}
   \left.\pdr{\tp}{\tx}\right|_{\tx=0}  = \dfrac{2\tJ}{1 + \ph(3 - 2 \yh)} \equiv W
   \label{eq:dpdx0}
\end{equation}
% Eliminating $\pdra{\tp}{\tx}$ from Eq.\eqref{eq:dydx0} by means of \eqref{eq:dpdx0}, we find
% \begin{equation}
%    \left. \pdr{y}{\tx}\right|_{\tx=0}
%            =  -  \dfrac{\yh\bigl(1 + \ph\bigr)2\tJ}{\ph\left(1 + \ph(3 - 2 \yh)\right)} - (1 + K)\tJ
%            \equiv U
%        \label{eq:dydx0a}
% \end{equation}

\section{Conservation law and approximate analytical solution}

% Integrating Eqs.\eqref{eq:dgm_hyd2}, \eqref{eq:dgm_sumd3} from 0 to $\tx$, we arrive at
Since $\tN_h = \tJ$, Eqs.\eqref{eq:dgm_hyd}, \eqref{eq:dgm_sumd} can be written as
\begin{align}
   & \pdr{(y\tp)}{\tx} + y\tp\pdr{\tp}{\tx} = {} - (1 + K) \tJ
   \label{eq:hyd4} \\
   & \bigl(1 + \tp\,(3 - 2y)\bigr)\pdr{\tp}{\tx} = 2 \tJ
   \label{eq:sumd4}
\end{align}
With the constant right sides, Eqs.\eqref{eq:hyd4}, \eqref{eq:sumd4} automatically satisfy
the mass transport Eqs.\eqref{eq:dgm_hyd2}, \eqref{eq:dgm_sumd3}.
Multiplying Eq.\eqref{eq:hyd4} by 2 and summing with Eq.\eqref{eq:sumd4}, we get
\begin{equation}
   2\pdr{(y\tp)}{\tx} + \bigl(1 + 3\tp\bigr)\pdr{\tp}{\tx} = - 2 K \tJ
   \label{eq:sum2}
\end{equation}
Rewriting this equation as
\begin{equation}
   2\pdr{(y\tp)}{\tx} + \pdr{\tp}{\tx} + \dfrac{3}{2}\pdr{\left(\tp^2\right)}{\tx} =  - 2 K \tJ
   \label{eq:sum2b}
\end{equation}
we see that it can be integrated from 0 to $\tx$ yielding a conservation law:
\begin{equation}
   2(y\tp - y_c\tp_c) + \left(\tp - \tp_c\right)
       + \dfrac{3}{2}\left(\tp^2 - \tp_c^2\right) = - 2 K\tJ \,\tx
   \label{eq:law}
\end{equation}
% Possible applications of Eq.\eqref{eq:law} are discussed below.

An analogue of Eq.\eqref{eq:law} can be derived for the two--component  CO-CO$_2$
system. Eq.\eqref{eq:hyd4} does not change, while Eq.\eqref{eq:dgm_sum}
in the dimensionless form reads
\begin{equation}
   \left(1 + \tp\,\bigl(\Kw - y\bigl(\Kw - 1\bigr) \bigr)\right)\pdr{\tp}{\tx}
      = - \left(1 - \Kw\right) \tJ
   \label{eq:dsum}
\end{equation}
Multiplying Eq.\eqref{eq:hyd4} by $(\Kw - 1)$ and summing with Eq.\eqref{eq:dsum} we get
\begin{equation}
   (\Kw - 1)\pdr{(y\tp)}{\tx} + \bigl(1 + Q\tp\bigr)\pdr{\tp}{\tx} = - (\Kw - 1) K \tJ
   \label{eq:dsum2}
\end{equation}
Integrating Eq.\eqref{eq:dsum2} from 0 to $\tx$ we find
\begin{multline}
   \bigl(Q - 1\bigr)(y\tp - y_c\tp_c) + \left(\tp - \tp_c\right)
       + \dfrac{Q}{2}\left(\tp^2 - \tp_c^2\right) \\
        = {} - \bigl(Q - 1\bigr) K\tJ \,\tx
   \label{eq:lawCO}
\end{multline}
where $Q = \sqrt{44/28}$ for the CO-CO$_2$ pair.

A simple analytical formula for the hydrogen molar
fraction through the ASL depth can be obtained from
the conservation law as follows.
The pressure $\tp$ slightly increases along $\tx$, while the hydrogen molar fraction $y$ decreases.
The variation of the product $\tp y$ along $\tx$ is not large and hence
the factor $\bigl(1 + \tp\,(3 - 2y)\bigr)$ in Eq.\eqref{eq:sumd4}
does not change much along $\tx$.
Thus, the deviation of $\pdra{\tp}{\tx}$ in Eq.\eqref{eq:sumd4} from a constant value is expected to be small.
From Eq.\eqref{eq:dpdx0} it follows that a reasonably
good approximation for $\tp(\tx)$ is a linear function
\begin{equation}
    \tp \simeq \ph + W\tx
    \label{eq:tp0}
\end{equation}
Solving Eq.\eqref{eq:law} for $y$ and substituting $\tp$, Eq.\eqref{eq:tp0}, into the resulting
equation, we get
\begin{multline}
   y = \dfrac{1}{\tp_c + W\tx}\biggl(y_c\tp_c - \dfrac{3}{4} W^2\tx^2 \\
   - \left(\dfrac{(3\tp_c + 1)}{2} W + K\tJ\right)\tx\biggr)
   \label{eq:ysol}
\end{multline}

At $\tx=1$, multiplying Eqs.\eqref{eq:ysol} and \eqref{eq:tp0} we get
the hydrogen molar concentration $c_{a} = p_* \tp_a y_a/(RT)$ at the active layer:
\begin{multline}
   c_{a} = \dfrac{p_*}{ RT}\left(y_c\tp_c - \dfrac{3}{4} W^2
      - \left(\dfrac{(3\tp_c + 1)}{2} W + K\tJ\right)\right), \\
       \text{mol~m$^{-3}$}
    \label{eq:ch}
\end{multline}
Similar results for CO-CO$_2$ mixture can be easily derived from
Eqs.\eqref{eq:lawCO}, \eqref{eq:tp0}.
Cell performance models include calculation of the anode concentration polarization, which requires
an accurate value of the hydrogen concentration $c_a$ in the anode active layer. Eq.\eqref{eq:ch}
gives a more accurate value of this parameter.
The anode transport overpotential $\eta_{tra}$ can be estimated using the Nernst equation
    \begin{equation}
       \eta_{tra} = \dfrac{RT}{2 F} \lnl{\dfrac{c_{c} c_{w,a}}{c_{a} c_{w,c}}}
                  =   \dfrac{RT}{2 F} \lnl{\dfrac{c_{c} (c_{t,a} - c_{a})}{c_{a} (c_{t,c} - c_{c})}}
    \end{equation}
    where the subscripts $c, a$ denote the channel and the active layer,
    and the subscript $t$ denotes the total molar concentration. In this equation, $c_{a}$ is given by Eq.\eqref{eq:ch},
    $c_{t,a} = p_a/(RT)$, where $p_a$ is given by Eq.\eqref{eq:tp0} with $\tx=1$
    and the other parameters are those in the channel. Note that Eq.(30) gives a rough estimate for $\eta_{tra}$,
    since the Nernst equation is not strictly valid in non--equilibrium conditions.

\section{Results and discussion}

The parameters used in the calculations below are collected in Table~\ref{tab:parms}.
The transport coefficients have been calculated as
\begin{equation}
   \begin{split}
      & B_0 = \dfrac{\lambda d^2}{32}, \quad \text{Ref.\cite{Bertei_15}} \\
      & D_{K,h} = \dfrac{\lambda d}{3} \sqrt{\dfrac{8 R T}{\pi M_{h}}} \\
      & D_m = \lambda D_m^{free}
   \end{split}
   \label{eq:parms}
\end{equation}
where $\lambda$ is the porosity/tortuosity ratio, $d$ is the mean pore diameter (Table~\ref{tab:parms}).
The numerical solution to Eqs.\eqref{eq:hyd4}, \eqref{eq:sumd4} was obtained using the standard Python
BVP solver {\em solve\_bvp}.

\begin{table}
\small
\begin{tabular}{|l|c|c|}
\hline
Cell temperature, K                  & $T$   &  273 + 800   \\
Pressure in the channel, Pa          & $p_c$ &  $10^5$      \\
Current density, A~m$^{-2}$          & $J$   &  $10^4$      \\
Anode thickness, m                   & $L$   &  $10^{-3}$   \\
Mean pore diameter, m                & $d$   &  $10^{-6}$   \\
Porosity/tortuosity ratio            & $\lambda$  & 0.033, Ref.\cite{Bao_07} \\
Hydrogen viscosity at 800 $^\circ$C, Pa~s    & $\mu$ & $2\cdot 10^{-5}$ \\
Free binary molecular diff. m$^2$~s$^{-1}$   & $D_m^{free}$ &  $8.154 \cdot 10^{-4}$, Ref.\cite{Bao_07}  \\
Anode gas composition                &       &  85\%H$_2$ + 15\%H$_2$O \\
% Anode mean pore radius, $\mu$m  &  $r$ &  1.3 \\
% Effective binary diffusion coefficient     &       & \\
%      in H$_2$-H$_2$O mixture, cm$^2$~s$^{-1}$   &  $D_m$ & 0.07, Ref.\cite{He_10} \\
\hline
\end{tabular}
\caption{Cell parameters used in calculations.
   }
\label{tab:parms}
\end{table}

\begin{figure}
    \begin{center}
        \includegraphics[scale=0.5]{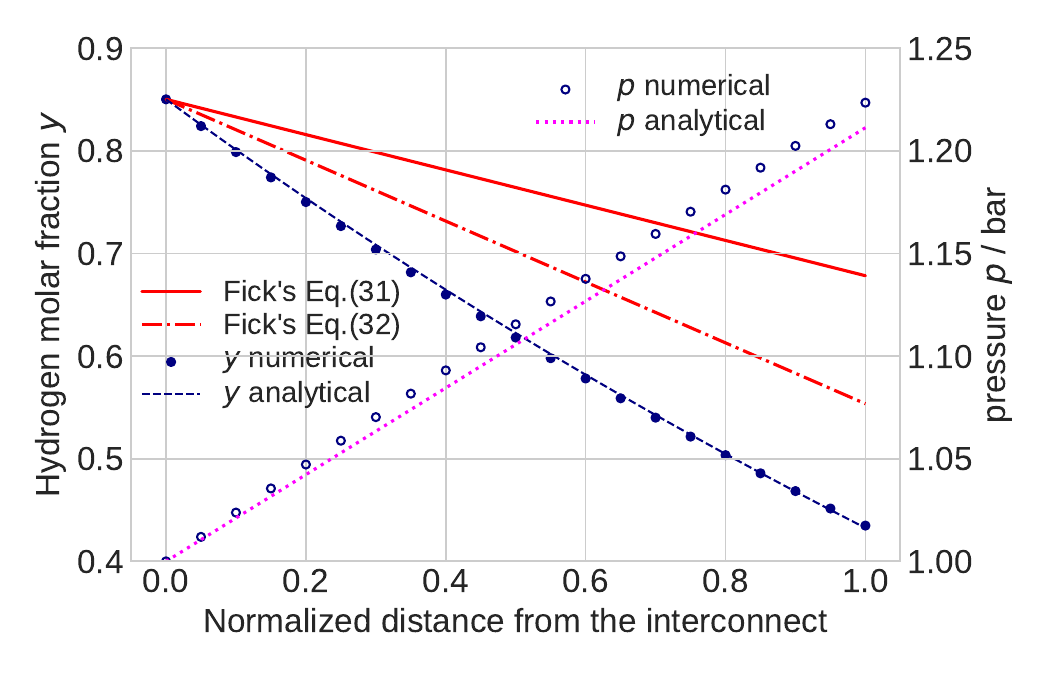}
        \caption{The shapes of hydrogen molar fraction through the one--mm thick Ni--YSZ cermet
        anode calculated using the Fick's law Eq.\eqref{eq:dyFick} (solid line),
        Eq.\eqref{eq:dyFick2} (dash-dotted line), from the
        numerical solution to the DGM model Eqs.\eqref{eq:hyd4}, \eqref{eq:sumd4} (solid points),
        and using the analytical solution Eq.\eqref{eq:ysol} (dashed line).
        Open circles show numerical pressure from the DGM model (right axis),
        dotted line is the approximate linear formula  for $p(\tx)$, Eq.\eqref{eq:tp0}.
        }
        \label{fig:yx}
    \end{center}
\end{figure}

Setting in Eq.\eqref{eq:law} $\tp = \tp_c$
we obtain the linear hydrogen molar fraction shape (Fick's law)
corresponding to zero pressure gradient in the ASL:
\begin{equation}
   y = y_c - \dfrac{ R T L J}{2 F D_m p_c}\,\dfrac{x}{L}
   \label{eq:dyFick}
\end{equation}
In the literature, another version of the Fick's law, which follows from the SMKM
has been used\cite{Suwa_03,Ma_21}:
\begin{equation}
   y = y_c - \dfrac{R T L J}{2 F p_c}\left(\dfrac{1}{D_m} + \dfrac{1}{D_{K,h}} \right)\dfrac{x}{L}
   \label{eq:dyFick2}
\end{equation}
Figure~\ref{fig:yx} shows
the linear shapes of $y$ from the Fick's law, Eq.\eqref{eq:dyFick} and Eq.\eqref{eq:dyFick2},
the exact numerical solution to the problem \eqref{eq:hyd4}, \eqref{eq:sumd4}, and
the approximate analytical Eq.\eqref{eq:ysol}.  The numerical and
analytical, Eq.\eqref{eq:tp0}, shapes of pressure are also shown.

Both Fick's equations, Eqs.\eqref{eq:dyFick},\eqref{eq:dyFick2} quite significantly
overestimate the exact hydrogen
molar fraction (Figure~\ref{fig:yx}). Physically, Knudsen diffusion
leads to formation of the positive pressure gradient (Figure~\ref{fig:yx}),
which retards hydrogen diffusion towards the active (reaction) zone, while
Fick's equations ignore this effect. Furthermore,
the Fick's law is used in literature together
with the constant pressure assumption, so the error in the calculated hydrogen
molar concentration $c = y p / (RT)$ is even larger than the error in $y$ itself shown in Figure~\ref{fig:yx}.

The analytical shape
of hydrogen molar fraction, Eq.\eqref{eq:ysol}, approximates the numerical result very well
(Figure~\ref{fig:yx}). The linear shape of pressure, Eq.\eqref{eq:tp0}, is somewhat less
accurate, though the maximal error at the ASL/active layer interface is about 1\% only
(Figure~\ref{fig:yx}). It is worth noting that for lower currents, the agreement of analytical
and numerical pressure shapes is better.
Eq.\eqref{eq:ch} could, thus, be recommended for calculation of concentration polarization
in SOFC anode instead of a widely used Fick's law.

The Fick's law, Eq.\eqref{eq:dyFick} results from the conservation law
under the condition $\pdra{\tp}{\tx} = 0$
and it shows that $y$ is independent of the Knudsen diffusivity.
Therefore, in the case of H$_2$-H$_2$O mixture,
the Stefan--Maxwell--Knudsen model, Eq.\eqref{eq:SMK_base}, and in particular
Eq.\eqref{eq:dyFick2}, is contradictory.
Indeed, from Eq.\eqref{eq:dyFick}
it follows that to neglect pressure gradient in Eq.\eqref{eq:dgm_base1},
{\em both} $\pdra{p}{x}$ and Knudsen $N_k/D_{K,k}$  terms must be omitted,
rather than just the term $\pdra{p}{x}$.

From Eq.\eqref{eq:tp0} an accurate condition for neglecting the pressure
gradient term can be derived. The pressure variation through the ASL is small
if the second term in Eq.\eqref{eq:tp0} at $\tx=1$ is much less than $\ph$.
Thus, the required condition is $W / \ph \ll 1$ and from Eq.\eqref{eq:dpdx0} we find
\begin{equation}
  \dfrac{2\tJ}{(1 + \ph(3 - 2 \yh))\ph} \ll 1
  \label{eq:tcond}
\end{equation}
In dimensional form Eq.\eqref{eq:tcond} reads
\begin{equation}
  \dfrac{J R T L}{F D_{K,h}\left(1 + \dfrac{p_c B_0(3 - 2 y_c)}{\mu D_{K,h}}\right)p_c} \ll 1
  \label{eq:cond}
\end{equation}
The hydraulic permeability $B_0$ is quadratic in the pore diameter $d$,
while the hydrogen Knudsen diffusivity $D_{K,h}$ is linear in $d$, Eqs.\eqref{eq:parms}.
Thus, for sufficiently small $d$,  the second term in the denominator of Eq.\eqref{eq:cond}
can be neglected and this equation simplifies to
\begin{equation}
  \dfrac{J R T L}{F D_{K,h} p_c} \ll 1
  \label{eq:cond1}
\end{equation}
Eq.\eqref{eq:cond1} does not contain $B_0$, i.e., the only process responsible
for the pressure gradient buildup is the Knudsen diffusion. For small $d$, $D_{K,h}$
is also small and Eq.\eqref{eq:cond1} is satisfied if the current density $J$
is sufficiently small.
% The exact estimate for the current can be calculated from Eq.\eqref{eq:cond1}.

In the opposite limit of large pore diameter, the unit in the denominator of Eq.\eqref{eq:cond}
can be neglected and this equation simplifies to
\begin{equation}
  \dfrac{J R T L \mu}{F B_0(3 - 2 y_c)p_c^2} \ll 1
  \label{eq:cond2}
\end{equation}
As expected, Eq.\eqref{eq:cond2} is independent of the Knudsen diffusivity,
i.e. the pressure gradient is developed by the finite hydraulic permeability
of the anode. At high cell current and/or low pressure in the anode channel
Eq.\eqref{eq:cond2} may be violated.

The general condition for the smallness of the pressure gradient is given
by Eq.\eqref{eq:cond}. With the data in Table~\ref{tab:parms}, the left side
of Eq.\eqref{eq:cond} is $0.21$ (Figure~\ref{fig:yx}), showing that the pressure
gradient term is quite significant. However, for the four times larger mean pore diameter,
the left side of Eq.\eqref{eq:cond} decreases to $0.036$, indicating that the pressure
variation through the ASL is small.  In this case, the solutions derived above
are of course valid, though the simpler Eq.\eqref{eq:dyFick2} is equally accurate.

Fu \etal\cite{Fu_15} reported detailed numerical analysis of the DGM solutions
for the SOFC anode. They showed that in the electrode with sufficiently large pore
diameter, the pressure gradient term can be neglected and isobaric DGM, which
is equivalent to the SMKM above can be used.  In their experiments, the anode
thickness was 1.5 mm, the estimated anode porosity/tortuosity ratio was 0.1
(Figure 11\cite{Fu_15}) and the mean pore diameter was around 2 $\mu$m.
The anode hydraulic permeability $B_0$ was not specified in\cite{Fu_15} and for the estimate
we use Eq.\eqref{eq:parms}. The remaining parameters, including the cell current density
are taken from Table~\ref{tab:parms}. With this set of parameters the left side of
Eq.\eqref{eq:cond} is  $0.048$, i.e., the pressure at the active layer exceeds
the pressure in the channel by 5\% only. Under these conditions, in the case of
H$_2$--H$_2$O mixture the isobaric approximation
works quite well, as reported in\cite{Fu_15}. Note that Fu \etal\cite{Fu_15} considered
also a triple gas mixture H$_2$--H$_2$O--N$_2$ and due to the presence of N$_2$
Eq.\eqref{eq:cond} may underestimate the pressure gradient. Calculations of Fu \etal show
that the nitrogen molar fraction decreases towards the active layer, i.e.
the diffusion flux of nitrogen is directed towards the active layer.
Thus, to keep the total nitrogen flux at zero,  a higher $\pdra{p}{x}$  is required
to support a counterflow of nitrogen towards the anode channel.

Generally, if the anode transport parameters are available, estimation with Eq.\eqref{eq:cond}
allows the appropriate approximation for the hydrogen molar fraction in the electrode
to be selected. However, it is more safe to use the equations from the previous section,
which are valid for any set of anode transport and operating parameters.

The conservation law, Eq.\eqref{eq:law}, provides several opportunities.
Setting in Eq.\eqref{eq:law} $\tx=1$, we get a relation between parameters
in the channel and at the ASL/active layer interface:
\begin{equation}
   2(y_a\tp_a - y_c\tp_c) + \left(\tp_a - \tp_c\right)
       + \dfrac{3}{2}\left(\tp_a^2 - \tp_c^2\right) = - 2 K\tJ
   \label{eq:solsumx1}
\end{equation}
where the subscript $a$ marks the values at $\tx=1$.
In dimensional form Eq.\eqref{eq:solsumx1} reads
\begin{multline}
   2(y_a p_a - y_c p_c) + p_a - p_c
       + \dfrac{3}{2}\left(p_a^2 - p_c^2\right)\dfrac{B_0}{\mu D_{K,h}} \\
           = - \dfrac{R T L J}{F D_m}
   \label{eq:dsolsumx1}
\end{multline}
From a practical perspective, by measuring the pressure $p_a$ one can calculate
the hydrogen molar fraction $y_a$ using Eq.\eqref{eq:dsolsumx1}.
Other useful options arise in the case of limiting current density:
setting in Eq.\eqref{eq:dsolsumx1} $y_a=0$, we get the relation between
$y_c$, $p_c$, $p_a$ and the system transport parameters:
\begin{equation}
   - 2y_c p_c  + p_a - p_c
       + \dfrac{3}{2}\left(p_a^2 - p_c^2\right)\dfrac{B_0}{\mu D_{K,h}}
           = - \dfrac{R T L J}{F D_m}
   \label{eq:dsolsumx2}
\end{equation}
Eq.\eqref{eq:dsolsumx2} allows direct calculation of the pressure $p_a$
at the active layer. On the other hand, by measuring $p_a$ in this regime
the relationship between $D_{K,h}$, $B_0$ and $D_m$ is obtained, so any one
of these three transport parameters can be estimated if the other two are known.
Measuring the pressure inside of the porous sandwich at high temperature is a challenging task~\cite{Nagata_93}.
However, it may be feasible in the future.

Finally, we note that the analytical shapes Eqs.\eqref{eq:tp0}, \eqref{eq:ysol}
can be used to develop a fast analytical model for the anode impedance.
This work is in progress and the results will be published elsewhere. Note also that the
hydrogen transport in the Solid Oxide Electrolysis Cells (SOECs) could be described by the same model.

\section{Conclusions}

An exact first integral (the conservation law) of the Dusty--Gas Model
for the two-component mixture transport in the SOFC anode support layer
is derived. Based on this result, an approximate analytical solutions
for the hydrogen molar fraction and total pressure shapes in the ASL are
obtained. Comparison with the numerical solution of the full system of
DGM equations shows the good quality of the approximate solutions.
A simple formula for the hydrogen molar concentration at the ASL/active
layer interface could be used for a more accurate calculation
of the concentration overpotential instead of the widely used Fick's law.
Several possibilities for measuring the electrode transport parameters
provided by the conservation law are discussed.

\newpage

% \bibliographystyle{unsrtnat}
% \bibliography{SOFC,Cell_08,Cell_0304,Cell_13,Cell_19,Cell,Impedance,MyPapers}

% \newpage

\section*{Nomenclature}

\small

\begin{tabular}{ll}
    $\tilde{}$   &  Marks dimensionless variables \\
    $B_0$        &  Hydraulic permeability, m$^2$, Eq.\eqref{eq:parms}       \\
    $c$          &  Hydrogen molar concentration, mol~m$^{-3}$           \\
    $c_a$        &  Hydrogen molar concentration \\
                 & at the active layer, mol~m$^{-3}$           \\
%     $\cref$      &  Reference oxygen concentration, mol~m$^{-3}$      \\
    $D_m$        &  Effective binary molecular diffusion coefficient    \\
                 &  in H$_2$--H$_2$O mixture, m$^2$~s$^{-1}$  \\
    $D_{K,h}$    &  Effective Knudsen diffusion coefficient \\
                 &  of hydrogen, m$^2$~s$^{-1}$  \\
    $D_{K,w}$    &  Effective Knudsen diffusion coefficient \\
                 &  of water, m$^2$~s$^{-1}$  \\
    $d$          &  Mean pore diameter, m                             \\
    $F$          &  Faraday constant, C~mol$^{-1}$                             \\
    $J$          &  Cell current density, A~m$^{-2}$     \\
    $K$          &  $ \equiv D_{K,h}/D_m$         \\
    $L$          &  Anode support layer thickness, m     \\
    $M_i$        &  Molecular weight of the $i$th component, kg~mol$^{-1}$ \\
    $N_i$        &  Molar flux of the $i$th component, mol~m$^{-2}$~s$^{-1}$        \\
    $p$          &  Pressure, Pa         \\
    $p_*$        &  Characteristic pressure, Pa, Eq.\eqref{eq:dless}    \\
    $Q$          &  $\equiv \sqrt{M_w/M_h} =  3$ \\
%     $q$          &  Dimensionless parameter, $0 \leq q \leq 1$      \\
    $R$          &  Gas constant, J~K$^{-1}$~mol$^{-1}$  \\
    $T$          &  Cell temperature, K \\
%     $U$          &  $\equiv \pdra{y}{\tx}|_{\tx=0}$, Eq.\eqref{eq:dydx0a}       \\
    $W$          &  $\equiv \pdra{\tp}{\tx}|_{\tx=0}$, Eq.\eqref{eq:dpdx0}       \\
    $x$          &  Coordinate through the anode support layer, m               \\
    $y$          &  Molar fraction of hydrogen                              \\
    $y_i$        &  Molar fraction of the $i$th component   \\[1em]
\end{tabular}

{\bf Subscripts:\\}

\begin{tabular}{ll}
    $*$       & Characteristic value \\
    $a$       & ASL/active layer interface \\
    $c$       & Channel/ASL interface     \\
    $h$       & Hydrogen \\
    $K$       & Knudsen diffusion \\
    $m$       & Molecular diffusion \\
    $t$       & Total molar concentration \\
    $w$       & Water             \\[1em]
\end{tabular}

% {\bf Superscripts:\\}
%
% \begin{tabular}{ll}
%     $0$      & Steady--state value \\
%     $1$      & Small--amplitude perturbation \\[1em]
% \end{tabular}

{\bf Greek:\\}

\begin{tabular}{ll}
    $\eta_{tra}$    & Anode transport overpotential \\
    $\lambda$       &  Porosity/tortuosity ratio  \\
    $\mu$           & Dynamic viscosity, Pa~s   \\
\end{tabular}

\newpage

\end{document}